\documentclass[11pt,a4paper]{article}
\pdfoutput=1
\usepackage{amssymb,amsmath,enumerate,mathrsfs,amsfonts}
\usepackage{graphicx,rotate,multicol}
\usepackage{multirow}
\usepackage{array}
\usepackage{jheppub}
\usepackage{blkarray}
\usepackage{multirow}
\usepackage{pdflscape}

\usepackage{dcolumn}
\usepackage{bm}
\usepackage{dcolumn}
\usepackage{textcomp}
\usepackage{float}
\usepackage{subfig}
\usepackage{hypcap}
\usepackage{adjustbox}
\usepackage{makecell}
\usepackage{color}
\usepackage{pifont}
\usepackage{appendix}
\usepackage{url}
\usepackage{cancel}
\allowdisplaybreaks[1]
\usepackage{import}
\usepackage{subfiles}
\usepackage{filecontents}
\usepackage{slashed}
\usepackage{lineno}
\usepackage{placeins}
\usepackage{booktabs}
\usepackage{soul}

\let\Oldsection\section
\renewcommand{\section}{\FloatBarrier\Oldsection}

\let\Oldsubsection\subsection
\renewcommand{\subsection}{\FloatBarrier\Oldsubsection}

\let\Oldsubsubsection\subsubsection
\renewcommand{\subsubsection}{\FloatBarrier\Oldsubsubsection}
\hypersetup{
  unicode=false,          
  pdftoolbar=true,        
 pdfmenubar=true,        
 pdffitwindow=true,     
 pdfstartview={FitH},    
 pdfsubject={GUT and Topological defects},   
 pdfnewwindow=true,      
 pdfcreator={RevTeX},
 colorlinks=true,       
 linkcolor=red,          
 citecolor=blue,        
 filecolor=black,      
 urlcolor=blue,           
  }

\usepackage{latexsym}
\usepackage{color}
\usepackage{bm}
\usepackage{prettyref}
\usepackage{comment}
\usepackage{numprint}
\usepackage[squaren]{SIunits}
\usepackage{array,booktabs,tabularx,fancybox}

\usepackage{slashed}

\usepackage{rotating}

\usepackage{float}

\usepackage{color}
\usepackage{multirow}
\long\def\rpl#1!!#2!!{\textcolor{red}{#1} \textcolor{blue}{#2}}
\long\def\commagda#1!!{\textcolor{green}{#1}}

\newcolumntype{Z}{>{\centering\arraybackslash}X}

\def \order(#1){{\cal O} \left(#1 \right)}

\makeatletter
\renewcommand*\env@matrix[1][*\c@MaxMatrixCols c]{%
	\hskip -\arraycolsep
	\let\@ifnextchar\new@ifnextchar
	\array{#1}}
\makeatother

\allowdisplaybreaks

\newtheorem{theorem}{Theorem}[section]

\newtheorem{cor}{Corollary}

\newcommand{\ket}[1]{|{#1}\rangle}
\newcommand{\N}[1]{\widetilde \nu}
\newcommand{\Nm}[1]{\ket{\widetilde \nu^{(m)}_{#1}}}
\newcommand{\Nf}[1]{\ket{\widetilde\nu^{(f)}_{#1}}}
\newcommand{\nuf}[1]{\ket{\nu^{(f)}_{#1}}}
\newcommand{\num}[1]{\ket{\nu^{(m)}_{#1}}}

\title{
New limits on neutrino non-unitary mixings \\based on prescribed singular values 
}
\author[a]{Wojciech Flieger,}
\author[a,b]{Janusz Gluza,}  
\author[a]{Kamil Porwit}

\affiliation[a]{Institute of Physics, University of Silesia, Katowice, Poland}
\affiliation[b]{Faculty of Science, University of Hradec Kr\'alov\'e, Czech Republic}
\emailAdd{woj.flieger@gmail.com}
\emailAdd{janusz.gluza@us.edu.pl}
\emailAdd{kamil.porwit@smcebi.edu.pl}

\abstract{
Singular values are used to construct physically admissible 3-dimensional mixing matrices characterized as contractions. Depending on the number of singular values strictly less than one, the space of the 3-dimensional mixing matrices can be split into four disjoint subsets, which accordingly corresponds to the minimal number of additional, non-standard neutrinos. We show in numerical analysis that taking into account present experimental precision and fits to different neutrino mass splitting schemes, it is not possible to distinguish, on the level of 3-dimensional mixing matrices, between two and three extra neutrino states. It means that in 3+2 and 3+3 neutrino mixing scenarios, using the so-called $\alpha$ parametrization, ranges of non-unitary mixings are the same. However, on the level of a complete unitary 3+1 neutrino mixing matrix, 
using the dilation procedure and the Cosine-Sine decomposition, we were able to shrink bounds for the "light-heavy" mixing matrix elements. For instance,
in the so-called seesaw mass scheme, a new upper limit on  $\vert U_{e4} \vert $ is about two times stringent than before and equals 0.021. For all considered mass schemes the lowest bounds are also obtained for all mixings, i.e. $\vert U_{e4} \vert , \vert U_{\mu 4} \vert , \vert U_{\tau 4} \vert .$ New results obtained in this work are based on analysis of neutrino mixing matrices obtained from the global fits at the 95\% CL.
}
\begin{document}
\maketitle

\section{Introduction}

The existence of additional neutrino flavors is one of the main posers in neutrino physics. Such particles can exist in nature and there are many theory driven experimental studies \cite{Sorel:2004,Karagiorgi:2009,Kopp:2011,Abazajian:2012,Dib:2019,Gariazzo2017,Gandhi2015,Gluza:2015goa,Gluza:2016qqv,Golling:2016gvc,Dube:2017jgo,Mangano:2018mur,Antusch:2018bgr, Deppisch:2015qwa}. Though new neutrino states may exist, in the Standard Model (SM) additional right-handed states do not couple directly with $W$ and $Z$ bosons, thus they are dubbed "sterile". However, they may influence the Standard Model physics, as they mix with "active" Standard Model left-handed states, which may lead to observable anomalies in theoretically predicted experimental results for the three neutrino flavor framework. Some reported anomalies can be found in Refs.
\cite{PhysRevC.83.054615, PhysRevD.83.073006, Serebrov2019, PhysRevC.73.045805, PhysRevC.83.065504}, though the reason may be different, connected with experimental setups or assumptions, as discussed for instance recently in \cite{Ioannisian:2019kse}.

Closely related to the problem of neutrino mixings is the issue of neutrino masses. Masses of sterile neutrinos are not limited so far and ranging from (sub-)eV to TeV, and higher to the Planck scale.
They may be very massive and explain masses of known light neutrinos by the seesaw mechanism \cite{Minkowski:1977,Mohapatra:1980}.  The heavy neutrinos provide an interesting connection
to Dark Matter, Baryogenesis via Leptogenesis, and feebly interaction dark sectors, also referred to as
'Neutrino Portal' \cite{Mohapatra:2005wg,Adhikari:2016bei,Abada:2019zxq,Abada:2019lih}. There also exist some hints towards two additional sterile neutrinos with eV scale masses \cite{Peres:2000, Sorel:2003, Kopp:2011, Giunti:2011, Heeck:2012}. However, they contradict the latest muon neutrino disappearance results from {\texttt{MINOS}}/{\texttt{MINOS+}} and {\texttt{IceCube}} \cite{Adamson:2017,Aartsen:2017}. So, the situation is not clear concerning scales and the number of additional neutrino states in general, and further scrutinize studies are needed, both on experimental and theoretical sides. 

In this paper we investigate in detail our original idea \cite{Bielas:2017lok} on how the notions of singular values and contractions which are coming from the matrix theory, influence limits on neutrino mixing parameters within the standard PMNS mixing matrix framework \cite{Pontecorvo:1957qd, Maki:1962mu, Tanabashi:2018oca}, when additional neutrino states are added \cite{Bielas:2017lok,Bielas:2017exa}. The main conceptual ideas explored in the original work \cite{Bielas:2017lok} included: 

\begin{itemize}
\item The definition of the region of physically admissible mixing matrices.

\item The characterization of mixing matrices according to the number of singular values strictly less than one which translates to the minimal number of additional neutrinos.

\item The notion of the unitary dilation, a mathematical way to extend physical matrices to larger unitary matrices and the numerical methods of doing this based on the Cosine-Sine decomposition.
\end{itemize}

New neutrino states modify the PMNS matrix, it is no longer unitary and mixing between extended flavor and mass states is described by a matrix of dimension larger than three. This extended matrix should in general itself be unitary, meaning completeness of the "active-sterile" mixing is restored. Hence, studies of the violation of unitarity of the SM {PMNS} mixing matrix is suitable for finding a hint for new neutrino states. 
In this scenario mixing between an extended set of neutrino mass states $\{\num \alpha,\Nm \beta\}$ with flavor states $\{\nuf \alpha,\Nf \beta\}$ is described by
\begin{align}
\begin{pmatrix} { \nuf \alpha} \\ 
 \Nf \beta\end{pmatrix}  &=
 \begin{pmatrix} {{ {U}_{\texttt{PMNS}} }} & U_{lh} \\  U_{hl} & U_{hh}
  \end{pmatrix} 
\begin{pmatrix} {\num \alpha } \\ 
\Nm \beta\end{pmatrix} 
\equiv U  \begin{pmatrix} {{ \num \alpha }} \\ \Nm \beta\end{pmatrix}\;.
\label{ugen}
\end{align}
Such block structures of the unitary $U$ are present in many neutrino mass theories. 
Indices "$l$" and "$h$" in \eqref{ugen} stand for "light" and "heavy" as usually we expect extra neutrino species to be much heavier than known neutrinos; cf. the seesaw mechanism  \cite{Minkowski:1977sc,Yanagida:1979as,GellMann:1980vs,Mohapatra:1979ia}. 
They can also include light sterile neutrinos, which effectively decouple in weak interactions, but are light enough to be in quantum superposition with three SM active neutrino states and to take part in the neutrino oscillation phenomenon \cite{Capozzi:2016vac}. Equivalently, we could call these mixings "active-sterile" ones.

The observable part of the above is the transformation from mass $\num \alpha, \Nm \beta$ to SM flavor $\nuf \alpha$ states ($\nuf{} \equiv \vert \nu_e \rangle,\vert \nu_\mu \rangle,\vert \nu_\tau \rangle $) and reads 
\begin{eqnarray}
\nuf \alpha &=& 
\sum_{i=1}^3\underbrace{\left( {U}_{\texttt{PMNS}}  \right)_{\alpha i} {\num i}}_{\rm SM \;part}
+ 
\sum_{j=1}^{n_R}\underbrace{
{ \left(U_{lh}  \right)}_{\alpha j} \Nm j}
_{\rm BSM \; part}\;. \label{gen}
\end{eqnarray}  
If ${U}_{\texttt{PMNS}}$ is not unitary then  there necessarily is a light-heavy neutrino "coupling" and the mixing between sectors is nontrivial $U_{lh} \neq 0 \neq U_{hl}$.

From a theoretical point of view, it is easier to study such deviations by representing the 3-dimensional non-unitary mixing matrix $\cancel{U}_{\texttt{PMNS}}$ as a product of a unitary matrix and some other type of a matrix. Two decompositions used frequently in neutrino mixing studies are known as the $\eta$ and $\alpha$ parameterizations \cite{Antusch:2006vwa, FernandezMartinez:2007ms, Xing:2007zj, Xing:2011ur, Escrihuela:2015wra, Escrihuela:2016ube, Blennow:2016jkn}. The first decomposes a given matrix into a product of a unitary matrix and a Hermitian matrix while the second one decomposes a matrix into a product of a unitary matrix and a lower triangular matrix
\begin{eqnarray}
\cancel{U}_{\texttt{PMNS}}&=&(1-\eta)Y_{1}, \\
\cancel{U}_{\texttt{PMNS}}&=&(1-\alpha)Y_{2},\label{alpha}
\end{eqnarray}
where  $\eta, \alpha$ and unitary matrices  $Y_{1}$ and $Y_{2}$ are  restricted by experiments.

We approach the problem of non-unitarity of the neutrino mixing matrix differently and explore mathematical properties of matrices, and their consequences for neutrino mixing analysis. To describe the neutrino mixing and experimental data in a uniform way, singular values are perfect quantities, giving a possibility to identify regions of physically admissible $3\times3$ mixing matrices \cite{Bielas:2017lok}. Singular values restrict neutrino mixing ranges to a special class of matrices known as contractions which have the largest singular values $\sigma_1$ smaller than one,

\begin{equation}\sigma_{1}=||V||\leq 1,
\label{contr}
\end{equation}
where $\Vert  \cdot \Vert$ stands for an operator norm of the matrix $V$. Moreover, matrix norms can be used to measure deviation from unitarity on the different levels of the mixing matrix. More details about the matrix norms applied to neutrino mixings can be found in \cite{Bielas:2017lok, Flieger:2019nsb}. 
Further, the number of singular values less than one determines a minimal number of additional neutrinos. It is then tempting to use this characteristic and construct right from the beginning mixing matrices with prescribed singular values which slice physical space of mixings into disjoint regions. Such matrices can be easily compared with experimental results via the $\alpha$ parametrization. 
The goal of the present work is to establish for the first time how a strict decomposition due to singular values affects ranges of neutrino mixing matrix entries, and allowed minimal dimensions of extended by the dilation procedure mixing matrices. 

In particular, our work includes:
\begin{enumerate}

\item Introduction to neutrino physics the method of the inverse singular value problem. It allows to construct mixing matrices with encoded minimal number of additional neutrinos and to confront them with experimental data.

\item Analysis of the amount of the space for additional neutrinos based on deviations of singular values from unity.

\item A study of possible distinction between three scenarios with different number of additional neutrinos on the level of experimental data using singular values and corresponding division of the neutrino mixing space $\Omega$.

\item Detailed analysis of the 3+1 scenario including derivation of new analytic bounds for the "light-heavy" neutrino mixing as a function of the singular value $\sigma_3<1$.

\end{enumerate}

In the next section, we present a geometrical argument that motivates our work, methodology and experimental groundwork. Section 3 is devoted to the   analysis in which matrices with prescribed singular values are used to restrict current experimental bounds regarding the minimal number of additional neutrinos. 
In Section 4 a scenario with one additional neutrino is considered in more detail using the dilation procedure. New analytical bounds  for the "light-heavy" mixing sector are derived. The work is concluded with a summary and discussion of possible directions for further studies.

\section{Matrix theory: Non-unitary neutrino mixings and experimental data}

\subsection{Subsets of the region of physically admissible mixing matrices}
In \cite{Bielas:2017lok} a region $\Omega$ of physically admissible mixing matrices was defined as the convex hull spanned on $3 \times 3$ unitary $U_{\texttt{PMNS}}$ mixing matrices with parameters restricted by experiments. Thus all matrices in $\Omega$ must necessarily be contractions, i.e. matrices with a spectral norm less or equal to one, see (\ref{contr}). It has also been shown that singular values control the minimal dimension of possible extensions of the $3\times3$ matrix $\cancel{U}_{\texttt{PMNS}}$
to a complete unitary matrix of some BSM models, which means that the minimal number of additional neutrinos is not arbitrary but depends on singular values. A distinction between the minimal dimension of the unitary extension of matrices from the region $\Omega$ is encoded in the number of singular values strictly smaller than one. This fact allows to divide $\Omega$ into four disjoint subsets: $\Omega_{1}$, $\Omega_{2}$, $\Omega_{3}$ and $\Omega_{4}$ characterized as 

\begin{eqnarray}
     \Omega_{1}:&& \text{3+1 scenario: } \Sigma = \lbrace \sigma_{1}=1.0, \sigma_{2}=1.0, \sigma_3 < 1.0 \rbrace  , \label{eqv1} \\
     \Omega_{2}:&& \text{3+2 scenario: } \Sigma = \lbrace \sigma_{1}=1.0, \sigma_{2}<1.0, \sigma_3 < 1.0 \rbrace \label{eqv2} ,\\
    \Omega_{3}:&& \text{3+3 scenario: } \Sigma = \lbrace \sigma_{1}<1.0, \sigma_{2}<1.0, \sigma_3 < 1.0 \rbrace  ,\label{eqv3}\\
    \Omega_{4}:&& \text{PMNS scenario: } \Sigma = \lbrace \sigma_{1}=1, \sigma_{2}=1, \sigma_{3}=1 \rbrace \label{eqv4}.
\end{eqnarray}
\\
$\Omega_{4}$ contains only unitary matrices. Such an internal structure of $\Omega$ provides motivation for analysis of the neutrino mixing matrices with respect to the minimal number of additional neutrino states. 

\subsection{Mixing matrix with prescribed singular values \label{subsconstr}}
  
In general, finding a matrix with a specified set of singular values $\Omega_1,\Omega_2,\Omega_3$ defined in (\ref{eqv1})-(\ref{eqv3}) within all physically admissible mixing matrices in $\Omega$ is a cumbersome task. A solution is to construct right from the beginning mixing matrices with a given set of singular values. In mathematics such an approach is known as the inverse singular value problem \cite{Chu:2000:FRA:343559.343582} which is closely related to the inverse eigenvalue problem \cite{Chu:1998:IEP:278129.278130}.
As a basis for the construction of such mixing matrices, we can either use a general matrix structure or try to simplify the task by invoking a specific matrix decomposition. 
Here we consider the $\alpha$ parametrization (\ref{alpha}) and focus on the lower triangular matrix $\alpha$. To obtain lower triangular matrices with prescribed eigenvalues and singular values
the algorithm proposed in \cite{Kwong-Li:2001} is used.
Construction is based on the majorization relation between eigenvalues and singular values
\begin{equation}\label{eq:major}
\prod_{i=1}^{k} |\lambda_{i}| \leq \prod_{i=1}^{k} \sigma_{i},
\end{equation}
with equality when $k=n$, where $n$ is the dimension of the matrix. 
Moreover, it is known that for such matrices eigenvalues are situated on the main diagonal \cite{horn_johnson_2012}. Thus we are able to construct the following matrix which will be called the $\mathrm{A}$-matrix 
\begin{equation}
\mathrm{A} \equiv \left(
\begin{array}{ccc}
{\bf \textcolor{red}{a_{11}}} & 0 & 0 \\
a_{21} & {\bf \textcolor{red}{a_{22}}} & 0 \\
a_{31} & a_{32} & {\bf \textcolor{red}{a_{33}}}
\end{array}
\right),
\end{equation}
with singular values \textcolor{red}{$\boldsymbol{\sigma_{1}}, \boldsymbol{\sigma_{2}}, \boldsymbol{\sigma_{3}}$} (highlighted quantities are prescribed). 
In this way, imposing eigenvalues and singular values the minimal number of additional neutrinos can be classified. Thanks to the  construction, we are left just with three free parameters $a_{21}, a_{31}, a_{32}$.  

\subsection{Present limits on the $T$-matrix derived from  experimental data fits}
In Tab.~\ref{tab:expbounds} experimental bounds are given for the $T$-matrix defined as a distortion of unity by $\alpha$, $T=I-\alpha$, for different neutrino mass scenarios. We chose to work with the $T$-matrix as it multiplies directly  the unitary matrix in the decomposition \eqref{alpha}, giving the same singular values as $\cancel{U}_{\texttt{PMNS}}$.
\begin{table}[h!]
\begin{center}
\begin{tabular}{cccc}
\toprule
Entry\ & \texttt{(I)}: $m>\unit{}{EW}$ & \texttt{(II)}: $\Delta m^2\gtrsim \unit{100}{eV^2}$ & \texttt{(III)}: $\Delta m^2 \sim \unit{0.1 -1}{eV^2}$\\
\midrule
$T_{11}=1-\alpha_{11}$ & $0.99870 \div 1$ & $0.976 \div 1$ & $0.990 \div 1$ \\
$T_{22}=1-\alpha_{22}$ & $0.99978 \div 1$ & $0.978 \div 1$ & $0.986 \div 1$ \\
$T_{33}=1-\alpha_{33}$ & $0.99720 \div 1$ & $0.900 \div 1$ & $0.900 \div 1$\\
$ T_{21} = |\alpha_{21}| $ & $0.0 \div 0.00068$ & $0.0 \div 0.025$ & $0.0 \div 0.017$ \\
$ T_{31} = |\alpha_{31}|$ & $0.0 \div 0.00270$ & $0.0 \div 0.069$ & $0,0 \div 0.045$ \\
$ T_{32} = |\alpha_{32}| $ & $0.0 \div 0.00120$ & $0.0 \div 0.012$ & $0.0 \div 0.053$ \\
\bottomrule
\end{tabular}
\caption{Limits on the elements of the $T=I-\alpha$ matrix for different non-standard neutrino mass scenarios \texttt{(I)}-\texttt{(III)}. Limits 
on the $\alpha$ matrix elements $\alpha_{ij}$ are taken from \cite{Blennow:2016jkn} (with 95\% CL), which are obtained from the global fits \cite{Fernandez-Martinez:2016lgt} to the experiments \cite{Declais:1994su, Abe:2014gda, MINOS:2016viw, Astier:2001yj, Astier:2003gs}. Moreover, these data also include formalism connected with non-standard interactions \cite{Blennow:2016jkn}.} 
\label{tab:expbounds}
\end{center}
\end{table}

It becomes customary to fit experimental data for three classes of different mass splittings\footnote{There are other regions of masses considered in literature \cite{Cvetic:2018elt,Cvetic:2019rms,Chun:2019nwi,Liu:2019qfa} with estimations on the "active-sterile" mixing elements. Our analysis requires data for the complete $3\times 3$ mixing matrix, as in Tab.~\ref{tab:expbounds}.}. In the scheme \texttt{(I)} 
it is supposed that masses of sterile neutrinos are at the GeV level, and above.  We can call it the seesaw scheme. 
In this case, the bounds are the tightest which gives the least space for a non-unitary mixing among considered scenarios. It is due to a natural decoupling of heavy neutrino states, which is hard to avoid \cite{Gluza:2002vs}. Two other cases leave significantly more space for a possible non-unitary mixing.
The intermediate scale, scheme \texttt{(II)}, 
is an interesting region for many experiments like \texttt{MINOS}, \texttt{LSND}, \texttt{DUNE} or \texttt{SBN} \cite{Parke:2015goa, Adams:2013qkq, Dutta:2016glq, Kopp:2013vaa}. Finally, in the scheme \texttt{(III)}, an additional sterile neutrino is only slightly more massive than known three massive neutrinos. 

Heavy extra neutrinos modify $Z$ and $W$ boson couplings. Such modifications translate to very strong limits on the scheme \texttt{(I)}. Constraints for this scheme have been obtained from global fits in Ref.~\cite{Fernandez-Martinez:2016lgt}.
The limits on intermediate scale \texttt{(II)} were obtained from different experiments. 

The constrain on $\alpha_{11}$ has been obtained in the \texttt{BUGEY-3} experiment \cite{Declais:1994su} with comparable values obtained by \texttt{Daya-Bay} \cite{An:2016luf}. Values for $\alpha_{22}$ and $\alpha_{33}$ come from the \texttt{SK} atmospheric oscillation measurements \cite{Abe:2014gda}. The off-diagonal elements, $\alpha_{21}$ and $\alpha_{32}$, have been obtained by the \texttt{NOMAD}  experiment \cite{Astier:2003gs}, in agreement with \texttt{KARMEN} results \cite{Armbruster:2002mp}.
Constrains on the diagonal element $\alpha_{11}$ for scheme \texttt{(III)} are obtained by \texttt{BUGEY-3} \cite{Declais:1994su}, for $\alpha_{22}$ by \texttt{SK} \cite{Abe:2014gda}, and for $\alpha_{33}$ in Ref.~\cite{Abe:2014gda}.
Limits for not listed elements were obtained indirectly from the diagonal elements \cite{Blennow:2016jkn}.

Tab.~\ref{tab:expbounds} represents experimental restrictions imposed to our analysis. All elements of constructed $\mathrm{A}$-matrices must lie within these limits.\\

\section{Pinning down the $T$-matrix  with singular values: Numerical results} \label{sec3}

\subsection{Error estimation}\label{sec:error}
Singular values are continuous functions of the matrix elements \cite{horn_johnson_2012}. Therefore, small changes in matrix elements do not change drastically their values. Quantitatively such behavior can be described with the help of Weyl inequalities \cite{Weyl1912,rao}.  
Let us assume that a $V$ matrix which realizes some BSM scenario includes the error matrix $E$ of the form $V+E$.
	Using Weyl inequalities for decreasingly ordered pairs of singular values of $V$ and $V+E$, the following relation takes place  
	\begin{equation}
		|\sigma_i(V+E)-\sigma_i(V)|\leq||E|| . \label{weyl}
	\end{equation}
A precision for elements of the $T$-matrix in the massive case $m>\unit{}{EW}$ is $10^{-5}$ (Tab.~\ref{tab:expbounds}). In our analysis, we keep the same precision for all massive cases. This does not contradict experimental results since we still work within experimentally established intervals.
Thus, all entries of the error matrix can be taken as $E_{ij}\approx 0.00001$ and the uncertainty of the calculated singular values is bounded by $||E||=0.00003$.
	
\subsection{Numerical distinguishability and continuity of singular values}

To see under which circumstances we can definitely distinguish numerically between two matrices from two different subsets of the $\Omega$ in (\ref{eqv1})-(\ref{eqv3}), let us consider a simple two dimensional model of a lower triangular matrix.

We start from the following diagonal matrix
\begin{equation}
\left(
    \begin{array}{cc}
    0.99 & 0 \\
    0 & 0.99
    \end{array}
\right).
\end{equation}
This matrix has obviously two singular values strictly less than one, equal to $0.99$. Now we transform this matrix to the lower triangular form by adding to the position $(2,1)$ a parameter $\epsilon$:
\begin{equation}\label{eq:eps}
\left(
    \begin{array}{cc}
    0.99 & 0 \\
    \epsilon & 0.99
    \end{array}
\right).
\end{equation}
We will increase $\epsilon$ from 0 to 1 with a different precision (Tab.~\ref{tab:con}) to see which precision guarantee such distinction.

\begin{figure}[h!]
\begin{center}
\includegraphics[scale=0.9]{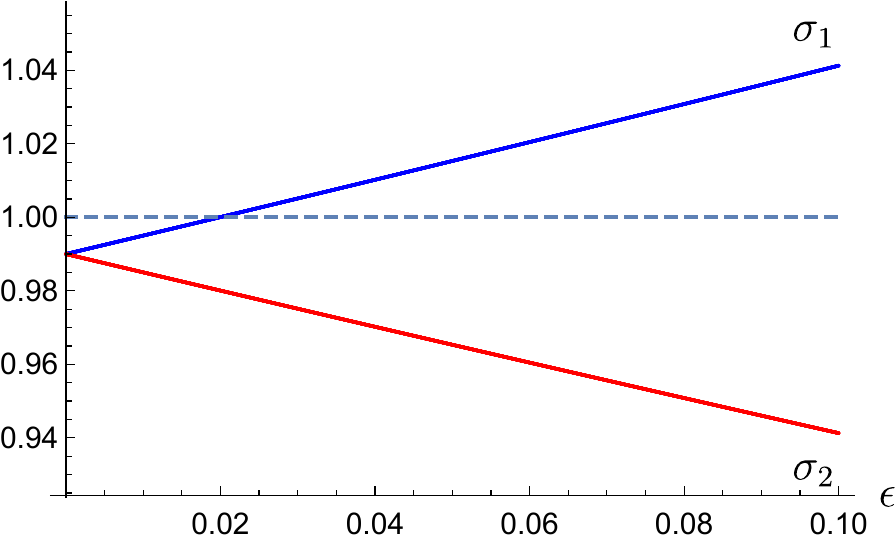}
\end{center}
\caption{
The behaviour of $\sigma_{1}$ and $\sigma_{2}$ as a function of the $\epsilon$ parameter which changes from 0 to 1. The continuity in changes of singular values with $\epsilon$ is evident. 
We are looking for the precision of the parameter $\epsilon$ for which $\sigma_{1}$ has within precision of calculations a common value with horizontal dashed line at 1, which represents the physical limit, see Table 2.
}
\label{fig1}
\end{figure}

Singular values will be fixed with the $3 \cdot 10^{-5}$ accuracy, as discussed in the previous subsection. Thus, we are interested in the precision of the $\epsilon$ which allows to distinguish singular values with this precision. In particular, we will determine when the first singular value can reach unambiguously the value one, i.e. $1\pm 3 \cdot 10^{-5}$. The results are gathered in the Tab.~\ref{tab:con}. We can see that the exact  determination of the singular value is possible for the precision of the $\epsilon$ at the level of $10^{-4}$, but it can not be achieved with the precision of the matrix element $\epsilon$ at the level of $10^{-3}$.
Further increase of  the $\epsilon$ precision
results in a larger number of solutions so, as long as we stay within allowed intervals, by adjusting a proper density of sifting, we are able to distinguish two singular values with imposed accuracy. This observation is a base for numerical calculations in this work, for details of numerical studies, see the Appendix.
 
\begin{table}[h!]
\begin{center}
\begin{tabular}{|c|c|}
\hline 
Initial singular value ($\sigma_{1}$) & 0.99 
\\
\hline 
Size of the $\epsilon$ step &  
{Number of matrices} \\
\hline 
0.001 & 0 
\\
0.0001 & 1 
\\
0.00001 & 11 
\\
\hline
\end{tabular}
\caption{A number of matrices that can reach the value of the first singular value ($\sigma_{1}$) equal to $1\pm 3\cdot 10^{-5}$ starting from 0.99. When the precision of  $\epsilon$ increases, it is more probably to find non-zero solutions.}
\label{tab:con}
\end{center}
\end{table} 
 
\subsection{Singular values as non-unitary mixing quantifiers in scenarios with N additional neutrinos}

We begin the analysis by testing how much space for additional states is available within current experimental limits using as quantifiers singular values. According to the division of $\Omega$ into subsets $\Omega_{1},\Omega_{2}$ and $\Omega_{3}$, it is sufficient to consider extensions by one, two or three neutrinos, and any larger extension is encoded in these three cases. To estimate the amount of mixing space for additional neutrinos we construct mixing matrices with prescribed singular values and decrease as much as possible the singular values that are not fixed to unity (\ref{eqv4})-(\ref{eqv3}).
The results are presented in Tab.~\ref{tab:complete}. 
All numerical results derived in this work are based on the the global fits to the experimental data \cite{Blennow:2016jkn} given with 95\% CL. We do not perform any additional fits to the experimental data and no additional statistical errors are induced. 
As discussed in Section \ref{sec:error}, the error connected with determination of singular values  is under control. It is connected with determination of  the interval matrix entries, in our case it is $3 \cdot 10^{-5}$ .

{\small{
\begin{table}[h!]
    \centering
		\begin{adjustbox}{max width=\textwidth}
			\begin{tabular}{cccc}
				\toprule
			$3+1$	& & & $\sigma_3$\\
				\midrule
				$m>\unit{}{EW}$ & & &  $0.9970\pm0.00003$\\
				$\Delta m^2\gtrsim \unit{100}{eV^2}$ & & & $0.900\pm0.00003$\\
				$\Delta m^2 \sim \unit{0.1 -1}{eV^2}$ & & & $0.889\pm0.00003$\\
				\midrule
			$3+2$	& & $\sigma_2$ & $\sigma_3$\\
				\midrule
				$m>\unit{}{EW}$ & & $0.9987\pm0.00003$ & $0.9986\pm0.00003$\\
				$\Delta m^2\gtrsim \unit{100}{eV^2}$ & & $0.976\pm0.00003$ & $0.975\pm0.00003$\\
				$\Delta m^2 \sim \unit{0.1 -1}{eV^2}$ & & $0.986\pm0.00003$	& $0.985\pm0.00003$\\
				\midrule
			$3+3$	& $\sigma_1$ & $\sigma_2$ & $\sigma_3$\\
				\midrule
				$m>\unit{}{EW}$ & $0.9998\pm0.00003$ & $0.9996\pm0.00003$ & $0.9996\pm0.00003$\\
				$\Delta m^2\gtrsim \unit{100}{eV^2}$ & $0.979\pm0.00003$ & $0.977\pm0.00003$ & $0.977\pm0.00003$\\
				$\Delta m^2 \sim \unit{0.1 -1}{eV^2}$ & $0.991\pm0.00003$ & $0.989\pm0.00003$ & $0.989\pm0.00003$\\
				\bottomrule
			\end{tabular}
		\end{adjustbox}
		\caption{The smallest possible values of $\sigma_i$ for different mass schemes in $3+N$ scenarios. The estimated error $3 \cdot 10^{-5}$ is written explicitly and is the same for all entries.}
    \label{tab:complete}
\end{table}
}}

The results gathered in Tab.~\ref{tab:complete} show that for each scenario there is space for additional neutrinos. From all massive schemes, the $m > \unit{}{EW}$ case contains the least space for extra neutrinos and in the scenario with 3 additional neutrinos, this space is limited most strongly. The most interesting situation occurs in the 3+1 case since the "free" singular value can be decreased the most. However, we should keep in mind that matrices from the subsets $V_{1}$ and $V_{2}$ can be also extended to higher dimensions, thus the space for two or three neutrinos should be treated in some sense cumulatively. In general, among all $3+N$ scenarios, the $\Delta m^2\gtrsim \unit{100}{eV^2}$ massive case leaves the most space for additional neutrinos.

\subsection{Bounds on the $\mathrm{A}$-matrix in scenarios with different number of additional neutrinos}

As the set of physically admissible mixing matrices splits into three disjoint subsets regarding the number of singular values strictly less than one, it is tempting to check for each disjoint subset, i.e. with one, two and three additional neutrinos, if we are able to shrink current experimental bounds for each case separately. 
This is done in two steps described  in the Appendix.

Experimental data in Tab.~\ref{tab:expbounds} for the $m>\unit{}{EW}$ case are given with the $10^{-5}$ precision while for remaining cases they are given with a precision of $10^{-3}$. 
However, in our analysis we keep consequently the $10^{-5}$ precision. 
It is possible as long as two matrices can be distinguished by their singular values with an imposed error, precision of individual elements is irrelevant, we fix it at the level of $10^{-5}$. 
The results of analysis are gathered in Tab.~\ref{tab:tab1}. Let us summarize them.

\begin{itemize}

\item For the 3+1 scenario, we observe shrinking in all elements of the $T$ matrix. 
Particularly, lower limits  for the element $(2,1)$ are non-zero and differ among massive scenarios \texttt{(I-III)}.
    
\item In the 3+1 scenario, the largest change of the lower bound is observed for the element {\color{red}{$(2,2)$}} in the $\Delta m^2\gtrsim \unit{100}{eV^2}$ case, while the largest change for the upper bound happens for the {\color{red}{$(2,1)$}} element and the same massive scenario \texttt{(II)}, see the red bold entries in the Tab.~\ref{tab:tab1}. 
     
\item We do not observe any differences between experimental values and our results in the 3+2 and 3+3 scenarios.
    
\end{itemize}

\begin{table}[h]
	\centering
		\begin{adjustbox}{max width=\textwidth}
			\begin{tabular}{ccccccc}
			\toprule
			& \texttt{(I)}: $m>\unit{}{EW}$ & \texttt{(II)}: $\Delta m^2\gtrsim \unit{100}{eV^2}$ & \texttt{(III)}: $\Delta m^2 \sim \unit{0.1 -1}{eV^2}$ \\
					\midrule
						$(1,1)$ & $0.99885 \div 0.99999$ & $0.97641 \div 0.99996$ & $0.99020 \div 0.99999$  \\
						Exp: & $0.99870 \div 1$ & $0.976 \div 1$ & $0.990 \div 1$ \\
						$(2,2)$ & $0.99980 \div 0.99999$ & $\textbf{\textcolor{red}{0.99331}} \div 0.99999$ & $0.98646 \div 0.99999$  \\
						Exp: & $0.99978 \div 1$ & $\textbf{\textcolor{red}{0.978}} \div 1$ & $0.986 \div 1$ \\
						$(3,3)$ & $0.99721 \div 0.99996$ & $0.90040 \div 0.99985$ & $0.90015 \div 0.99958$ \\
						Exp: & $0.99720 \div 1$ & $0.900 \div 1$ & $0.900 \div 1$ \\
						$(2,1)$ & $0.00001 \div 0.00062$ & $0.00031 \div \textbf{\textcolor{red}{0.02214}}$ & $0.00014 \div 0.01615$  \\
						Exp: & $0.0 \div 0.00068$ & $0.0 \div \textbf{\textcolor{red}{0.025}}$ & $0.0 \div 0.017$ \\
						$(3,1)$ & $0.00002 \div 0.00266$ & $0.00048 \div 0.06892$ & $0.00012 \div 0.04500$  \\
						Exp: & $0.0 \div 0.00270$ & $0.0 \div 0.069$ & $0.0 \div 0.045$ \\
						$(3,2)$ & $0.00008 \div 0.00113$ & $0.00052 - 0.01196$ & $0.00024 \div 0.05281$ \\
						Exp: & $0.0 \div 0.00120$ & $0.0 \div 0.012$ & $0.0 \div 0.053$ \\
					\bottomrule
			\end{tabular}
		\end{adjustbox}
\caption{Lower and upper bounds for elements of the matrix $T$ in the 3+1 scenario. For scenarios 3+2 and 3+3 we do not observe any changes of experimental limits given in Tab.~\ref{tab:expbounds}.}
	\label{tab:tab1}
\end{table}

\section{The 3+1 scenario}

\subsection{Spread of the elements for the smallest singular value}
As one may expect, there is a possibility to find more than one matrix with a given set of singular values. We analyze how big the spread of the elements of the $\mathrm{A}$-matrix is in the case of the 3+1 scenario when the smallest value of the third singular value is considered (Tab. \ref{tab:complete}).
\begin{small}
\begin{itemize}
\item \texttt{(I)}: $m>\unit{}{EW}, \Sigma=\lbrace 1, 1, 0.9970 \rbrace .$
\begin{eqnarray}\label{low1}
&&\vert \mathrm{A}_{0.9970} \vert =\\
&&\nonumber \\
&&\
\left(
\begin{array}{ccc}
0.999869 \div 0.999928 \ (\textcolor{red}{4.5\%}) & 0 & 0 \\
0.000192 \div 0.000263 \ (\textcolor{red}{10\%}) & 0.999869 \div 0.9999 \  (\textcolor{red}{14\%}) & 0 \\
0.000897 \div 0.001199 \ (\textcolor{red}{11\%}) & 0.001059 \div 0.0012 \ (\textcolor{red}{12\%}) & 0.9972 \div 0.997262 \ (\textcolor{red}{2\%})
\end{array}
\right).\nonumber 
\end{eqnarray}

\item \texttt{(II)}: $\Delta m^2\gtrsim \unit{100}{eV^2}, \Sigma=\lbrace 1, 1, 0.900 \rbrace .$
\begin{eqnarray}
&&\vert \mathrm{A}_{0.900} \vert = \\
&&\nonumber \\
&&\left(
\begin{array}{ccc}
0.999623 \div 0.999999 \ (\textcolor{red}{1.5\%}) & 0 & 0 \\
0.000002 \div 0.000753 \ (\textcolor{red}{3\%}) & 0.999623 \div 0.999999 \ (\textcolor{red}{2\%}) & 0 \\
0.000606 \div 0.011919 \ (\textcolor{red}{16\%}) & 0.000606 \div 0.011923 \ (\textcolor{red}{94\%}) & 0.900002 \div 0.900678 \ (\textcolor{red}{1\%})
\end{array}
\right).\nonumber
\label{low2}
\end{eqnarray}

\item \texttt{(III)}: $\Delta m^2 \sim \unit{0.1 -1}{eV^2}, \Sigma=\lbrace 1, 1, 0.889 \rbrace .$
\begin{equation}
\vert \mathrm{A}_{0.889} \vert =
\left(
\begin{array}{ccc}
0.994583 \div 0.995161 \ (\textcolor{red}{6\%}) & 0 & 0 \\
0.011947 \div 0.012565 \ (\textcolor{red}{4\%}) & 0.992566 \div 0.993158 \ (\textcolor{red}{4\%}) & 0 \\
0.04251 \div 0.045 \ (\textcolor{red}{5.5\%}) & 0.050903 \div 0.053 \ (\textcolor{red}{4\%}) & 0.9 \div 0.90047 \ (\textcolor{red}{0.5\%})
\end{array}
\right).
\label{low3}
\end{equation}
\end{itemize}
\end{small}

Results given in (\ref{low1})-(\ref{low3}) reveal that in the case of the smallest possible singular values the spread of elements is typically about 15\% or less of the experimental intervals given in Tab.~\ref{tab:expbounds}. The exact percentage is given in red in (\ref{low1})-(\ref{low3}). The exception is the element (3,2) for the intermediate massive case where almost the entire experimental interval is covered. This also shows that in the case of the one particular set of singular values experimental bounds can be narrowed substantially. 

\subsection{Dilation, a quest for complete mixing}

As discussed in \cite{Bielas:2017exa}, if the standard $3 \times 3$ neutrino mixing matrix $U_{\texttt{PMNS}}$ would be non-unitary, it should be a part of a larger unitary matrix. Contractions imply that also all matrices from the region $\Omega$ can be naturally expanded to a larger unitary matrix. Such a procedure is known as a unitary dilation and its reverse is known as compression. In the dilation case, our initial matrix $U_{\texttt{PMNS}}$ is located in the top-left corner of the extended complete unitary matrix $U$
\begin{equation}
U=
\left(
\begin{array}{cc}
U_{\texttt{PMNS}} & U_{lh} \\
U_{hl} & U_{hh}
\end{array}
\right),
\end{equation}
and $UU^{\dag}=U^{\dag}U=I$.

Numerically the unitary dilation can be done with use of the Cosine-Sine (CS) Decomposition \cite{Allen:2006}, which can be cast in the form of the following theorem.

\begin{theorem}
\label{th:2}
Let the unitary matrix $U \in M_{(n+m) \times (n+m)}$ be partitioned as
\begin{equation}
U=
\begin{blockarray}{ccc}
n & m & \\ 
\begin{block}{(cc)c}
U_{\texttt{PMNS}} & U_{lh} & \ n \\
U_{hl} & U_{hh} & \ m \\
\end{block}
\end{blockarray},
\end{equation}
If $m \geq n$, then there are unitary matrices $W_{1}, Q_{1} \in M_{n \times n}$ and unitary matrices $W_{2}, Q_{2} \in M_{m \times m}$ such that
\begin{equation}
\begin{split}
&\left(
\begin{array}{cc}
U_{\texttt{PMNS}} & U_{lh} \\
U_{hl} & U_{hh}
\end{array}
\right)= 
\left(
\begin{array}{cc}
W_{1} & 0 \\
0 & W_{2}
\end{array}
\right)
\left(
\begin{array}{c|cc}
C & -S & 0 \\ \hline
S & C & 0 \\
0 & 0 & I_{m-n}
\end{array}
\right)
\left(
\begin{array}{cc}
Q_{1}^{\dag} & 0 \\
0 & Q_{2}^{\dag}
\end{array}
\right),
\end{split}
\end{equation}
where $C \geq 0$ and $S \geq 0$ are diagonal matrices satisfying $C^{2} + S^{2}=I_{n}$.
\end{theorem}
If $n \geq m$ then it is possible to parametrize a unitary dilation of the smallest size.
\begin{cor}\label{cor:2}
The parametrization of the unitary dilation of the smallest size is given by
\begin{equation}
\begin{split}
&
\left(
\begin{array}{cc}
U_{\texttt{PMNS}} & U_{lh} \\
U_{hl} & U_{hh}\
\end{array}
\right)=
\left(
\begin{array}{cc}
W_{1} & 0 \\
0 & W_{2} 
\end{array}
\right)
\left(
\begin{array}{cc|c}
I_{r} & 0 & 0 \\
0 & C & -S \\ \hline
0 & S & C
\end{array}
\right)
\left(
\begin{array}{cc}
Q_{1}^{\dag} & 0 \\
0 & Q_{2}^{\dag}
\end{array}
\right),
\end{split}
\end{equation}
where $r=n-m$ is the number of singular values equal to 1 and $C=diag(\cos \theta_{1},...,\cos \theta_{m})$ with $\vert \cos \theta_{i} \vert <1$ for $i=1,...,m$.
\end{cor}
A knowledge of the experimental bounds for each of $3+N$, $N=1,2,3$ scenarios with help of the unitary dilation can be used to estimate limits for elements of the complete mixing matrix corresponding to the non-unitary mixing.

\subsection{New estimations for "active-sterile" mixings \label{subres}}
To estimate the "light-heavy" mixing sector in the case of one additional neutrino, the CS decomposition will be used.
In the 3+1 scenario only one singular value is different from unity and the $\mathrm{A}$-matrix has singular values $\sigma_{1}=1, \sigma_{2}=1, \sigma_{3}<1$. In such case the CS decomposition takes the form
\begin{equation}
\left(
\begin{array}{cc}
U_{\texttt{PMNS}} & U_{lh} \\
U_{hl} & U_{hh}\
\end{array}
\right)=\left(
\begin{array}{cc}
W_{1} & 0 \\
0 & W_{2} 
\end{array}
\right)
\left(
\begin{array}{ccc|c}
1 & 0 & 0 & 0\\
0 & 1 & 0 & 0 \\ 
0 & 0 & c & -s \\ \hline
0 & 0 & s & c
\end{array}
\right)
\left(
\begin{array}{cc}
Q_{1}^{\dag} & 0 \\
0 & Q_{2}^{\dag}
\end{array}
\right).
\end{equation}
For the "light-heavy" mixing sector we have
\begin{equation}
U_{lh}=W_{1}S_{12}Q_{2}^{\dag},    
\end{equation}
where $W_{1} \in \mathbb{C}^{3 \times 3}$ is unitary, $S_{12}=(0,0,-s)^{T}$ and $Q_{2} = e^{i \theta}, \theta \in (0, 2 \pi]$.
Parametrizing the matrix $W_{1}$ by Euler angles we get
\begin{equation}\label{eq:u12}
U_{lh}=-(-s_{12}e^{-i\theta_{13}},s_{23}c_{13},c_{23}c_{13})^{T}se^{-i\theta}\equiv -(w_{e3},w_{\mu 3},w_{\tau 3})^{T}se^{-i\theta}.
\end{equation}
To estimate the largest possible absolute values for the elements of the "light-heavy" sector, we get
\begin{equation}\label{eq:raw_estimation}
| s | = | \sqrt{1-c^{2}} | = | \sqrt{1-\sigma_{3}^{2}} |.
\end{equation}
For estimations, the smallest third singular value are taken from Tab.~\eqref{tab:complete}, which implies that the upper bound on the mixing is obtained. 
Thus, for each massive scenario we get
\begin{equation}
\begin{split}
m>\unit{}{EW}: \vert U_{i4} \vert <  0.08359 , \\
\Delta m^2\gtrsim \unit{100}{eV^2}: \vert U_{i4} \vert <  0.43795 , \\
\Delta m^2 \sim \unit{0.1 -1}{eV^2}: \vert U_{i4} \vert <  0.4579,
\end{split}    
\end{equation}
where $i=e, \mu, \tau$. 

This rough estimation can not distinguish mixings which involve different flavors. To make this distinction possible and to get as sharp bounds as possible, the exact form of the $W_{1}$ matrix is needed, which follows from the singular value decomposition of the lower triangular $\mathrm{A}$-matrices with prescribed singular values $\lbrace 1, 1, \sigma_{3} \rbrace$. $\sigma_{3}$ ranges from $\sigma_{3min}$ (Tab.~\ref{tab:complete}) up to maximally allowed values. In this case the "light-heavy" mixing is given by

\begin{equation}
\vert U_{i4} \vert= \vert w_{i3} \vert \cdot \vert \sqrt{1-\sigma_{3}^{2}} \vert  \label{eque4},  \quad i=e,\mu,\tau.
\end{equation}
For all combinations we have generated $10^{4}$ matrices which satisfy majorization relation \eqref{eq:major}, and with elements within the experimental bounds given in Tab.~\ref{tab:expbounds}. For each matrix, the singular value decomposition has been performed and the maximal and minimal absolute values of the $W_{1}$ matrix have been taken. 

\begin{figure}
\begin{center}
    \includegraphics[scale=0.5]{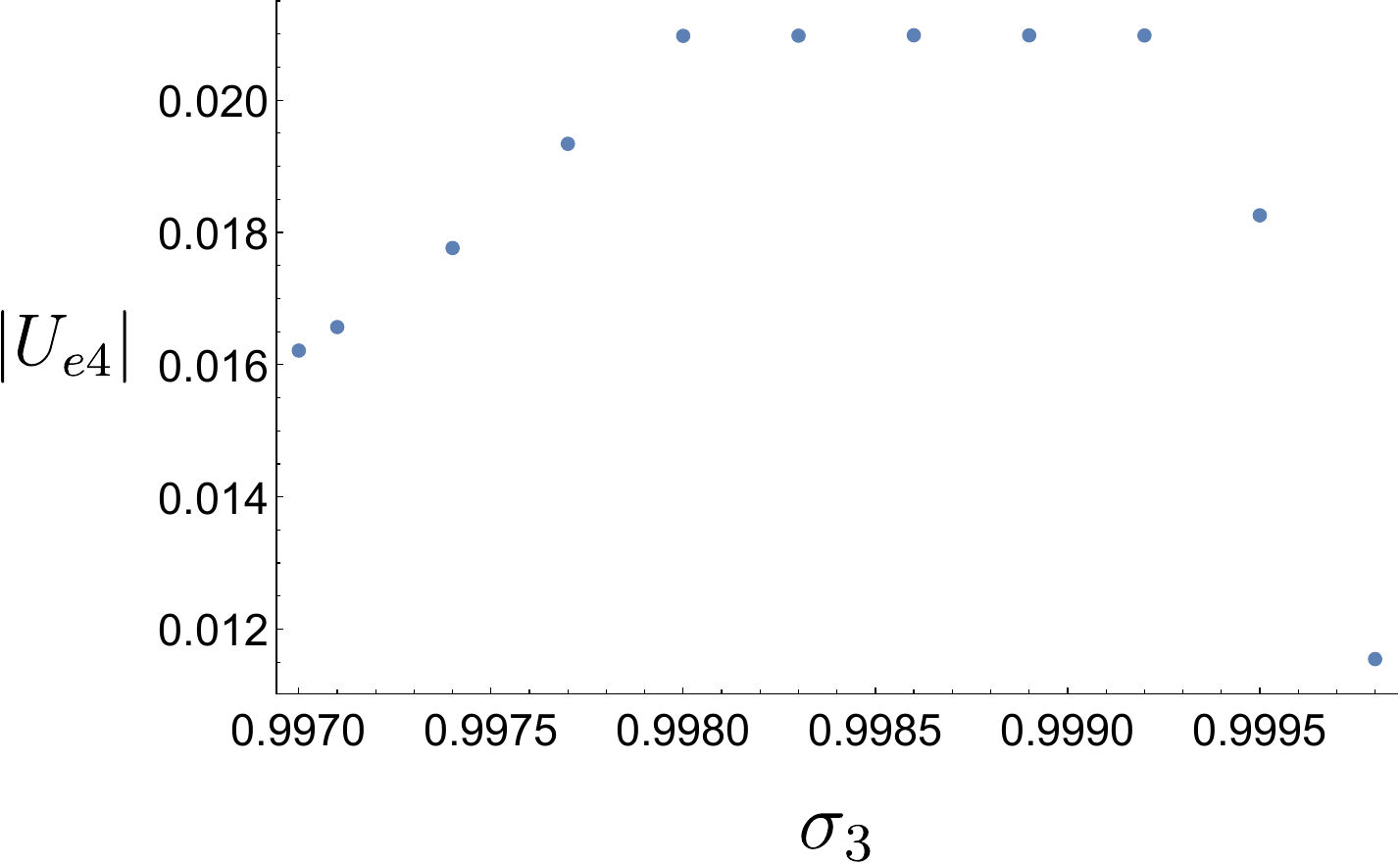} 
    \includegraphics[scale=0.5]{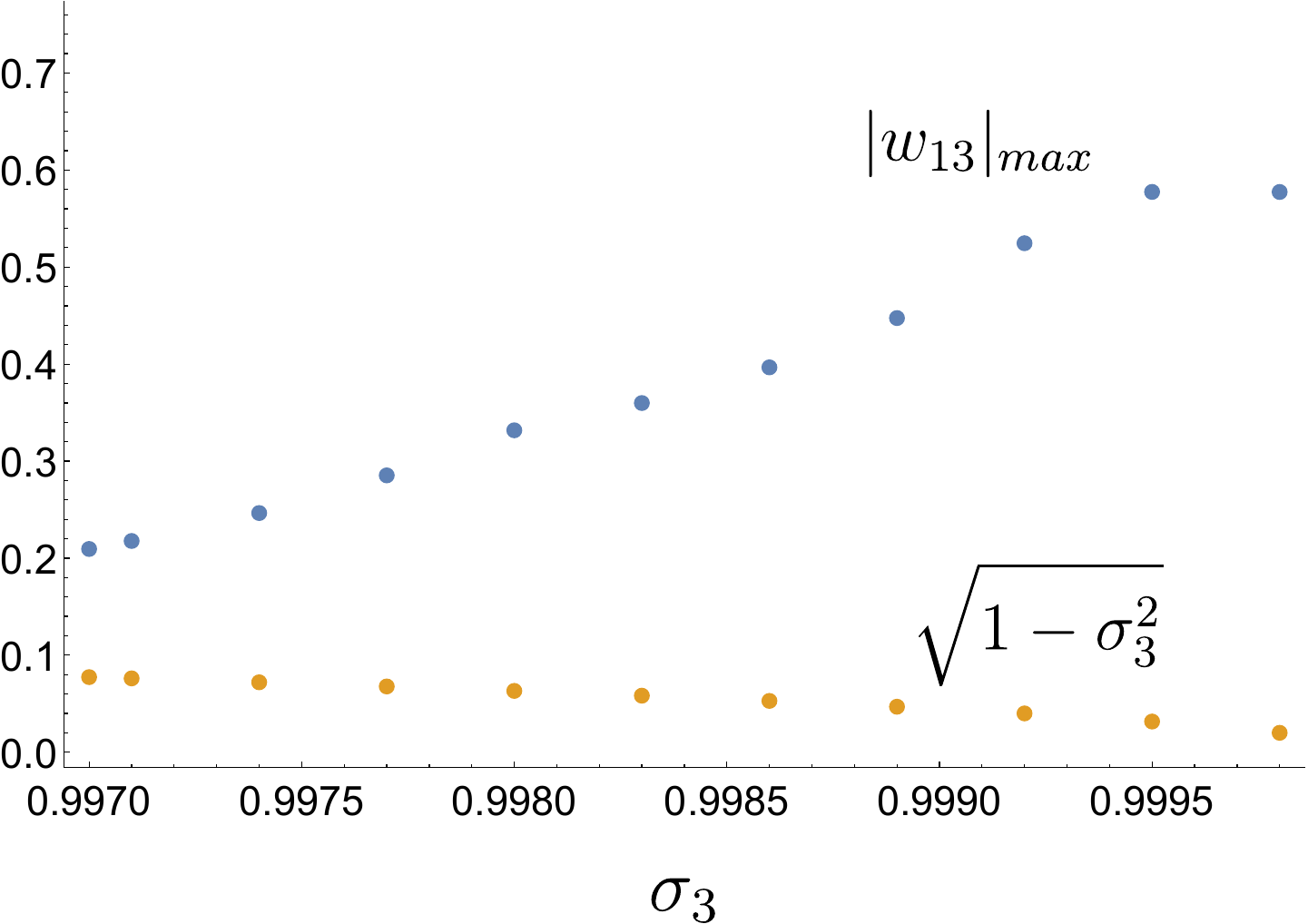}
    \end{center}
    \caption{Upper bounds for $\vert U_{e4} \vert$ as a function of $\sigma_{3}$ in the massive case \texttt{(I)} (left). $\vert U_{e4} \vert$ approaches the maximum for middle values of $\sigma_{3}$. On right, the behavior of particular constituents of $\vert U_{e4} \vert$, defined in (\ref{eque4}), as a function of  $\sigma_{3}$ is given.}
    \label{fig3}
\end{figure}

To justify the procedure described in this section, Fig.~\ref{fig3} shows the behaviour of the "light-heavy" mixing $\vert U_{e4} \vert$ as a function of $\sigma_{3}$  for  the massive case \texttt{(I)} in Tab.~\ref{tab:expbounds}, see also \cite{Flieger:2019grb}. It shows that the "light-heavy" mixing element depends continuously on $\sigma_{3}$ and its maximum is attained. In the $3+1$ scenario only the third column of the $W_{1}$ matrix is important. Thus performing the same steps as for our first estimation \eqref{eq:raw_estimation} and replacing $w_{e3},w_{\mu 3},w_{\tau 3}$ in \eqref{eque4} with obtained maximal and minimal values we get,
\begin{itemize}
\item \texttt{(I)}: $m>\unit{}{EW}.$
\begin{equation}
\begin{split}
&\vert U_{e4} \vert \in \left[0, 0.021 \right], \hspace{13pt}
\vert U_{\mu 4} \vert \in \left[0.00013, 0.021 \right], \quad
\vert U_{\tau 4} \vert \in \left[0.0115, 0.075 \right]. \\
&\vert U_{e4} \vert \leq 0.041 \ \text{\cite{deBlas:2013gla}}, \quad
\vert U_{\mu 4} \vert \leq 0.030 \ \text{\cite{deBlas:2013gla}}, \quad \quad \quad \quad
\vert U_{\tau 4} \vert \leq 0.087 \ \text{\cite{deBlas:2013gla}}.  \label{eqrescaseI}
\end{split}
\end{equation}
\item \texttt{(II)}: $\Delta m^2\gtrsim \unit{100}{eV^2}.$
\begin{equation}
\vert U_{e4} \vert \in \left[0, 0.082 \right], \quad
\vert U_{\mu 4} \vert \in \left[0.00052,0.099 \right], \quad
\vert U_{\tau 4} \vert \in \left[0.0365, 0.44\right].
\end{equation}
\item \texttt{(III)}: $\Delta m^2 \sim \unit{0.1 -1}{eV^2}.$
\begin{equation}
\begin{split}
&\vert U_{e4} \vert \in \left[0, 0.130 \right], \hspace{46.5pt}
\vert U_{\mu 4} \vert \in \left[0.00052, 0.167 \right], \hspace{18.7pt}
\vert U_{\tau 4} \vert \in \left[0.0365, 0.436 \right]. \\
&\vert U_{e4} \vert \in \left[0.114, 0.167 \right] \ \text{\cite{Gariazzo:2017fdh}}, \quad
\vert U_{\mu 4} \vert \in \left[0.0911, 0.148 \right]\ \text{\cite{Gariazzo:2017fdh}}, \quad
\vert U_{\tau 4} \vert \leq 0.361 \ \text{\cite{Dentler:2018sju}}.  \label{eqrescaseIII}
\end{split}
\end{equation}
\end{itemize}
 
Similarly to the results gathered in \cite{deBlas:2013gla,Gariazzo:2017fdh,Dentler:2018sju} (second rows in (\ref{eqrescaseI})-(\ref{eqrescaseIII})), the results obtained in this work  (first rows in (\ref{eqrescaseI})-(\ref{eqrescaseIII})) correspond to 95\% CL.
For the upper bound  in the case of the light sterile neutrino, scheme \texttt{(III)}, the limit obtained for $\vert U_{e4} \vert$ is tighter by about 20\% than those presented in the literature \cite{Gariazzo:2017fdh}.  
Similarly, for the seesaw like sterile neutrino, scheme \texttt{(I)}, the limit for $\vert U_{e4} \vert$ is about 50\% better than the current bound, and limits for $\vert U_{\mu 4} \vert$ and $\vert U_{\tau 4} \vert$ are respectively about 30\% and 10\%  better than present bounds \cite{deBlas:2013gla}. In \cite{Gariazzo:2017fdh} the lower bounds for $\vert U_{e4} \vert$ and $\vert U_{\mu 4} \vert$ for scheme \texttt{(III)} are better than our results, however, opposite is true for $\vert U_{\tau 4} \vert$.  We would like to stress that our method allows to obtain lower bounds for all massive cases.
	
\section{Conclusions and outlook}
BSM signals in neutrino physics can emerge as deviations from unitarity of the 3-dimensional PMNS mixing matrix. Here we invoked the notion of singular values which links in the elegant and simple way the experimental and theoretical knowledge on mixing matrices. We performed numerical analysis for singular values of experimental mixing matrix interval data in the $\alpha$ matrix representation, which reflects neutrino mixing distortion from unitarity. The method of construction of matrices with prescribed singular values was introduced which allows to study subsets of the region of admissible mixing matrices according to the minimal number of additional neutrinos. Therefore, emphasis has been put on the study of possibility of distinction among three different scenarios, i.e., 3+1, 3+2 and 3+3, on the level of the present experimental bounds. Firstly, we have estimated the amount of space available for additional neutrinos. Results show that within the present experimental limits, there is enough space for all number of additional neutrinos in the whole mass spectrum. However, in the case of 3 additional neutrinos with masses above the electro-weak scale this space is strongly limited. Analysis reveals that 3+2 and 3+3 scenarios are indistinguishable and the mixing space belonging to subsets $\Omega_{2}$ and $\Omega_{3}$ in (\ref{eqv2}),(\ref{eqv3}) covers the entire experimental intervals. More interesting is the 3+1 scheme where we observe shrunk of the experimental ranges for elements of the triangular $\mathrm{A}$-matrix. The most significant difference between obtained ranges and experimental values are given for elements (2,1) and (2,2), in the intermediate massive case (\texttt{II}). It is clear from continuity of singular values that the distinction between these three scenarios should emerge at some level of precision. 

Looking in more detail to the 3+1 scenario, the most interesting results are obtained in subsection~\ref{subres}, where the CS decomposition allowed to get lower and upper bounds on "light-heavy" neutrino mixings, i.e. the elements $\vert U_{e 4} \vert, \vert U_{\mu 4} \vert$ and $\vert U_{\tau 4} \vert$,  for all massive cases. It is worth noticing that also the limits on the tau neutrino mixing with a new neutrino state have been obtained.
In particular, we have improved substantially bounds for the seesaw scenario, e.g. the upper bound on $\vert U_{e 4} \vert$ is about 2 times better than in previous analysis. However, in the case of a sterile light neutrino (scheme \texttt{III}) the only tightened constraint has been obtained for the mixing between the electron neutrino and the fourth massive state. 

Matrix theory provides many useful tools to study the physics of particles mixing phenomenon. 
In the near future, we plan to extend the analysis especially to estimate the "light-heavy" mixing for the 3+2 and 3+3 scenarios. Moreover, we plan to study in detail connections between masses and neutrino mixings in the seesaw regime. In the long run, it is very important to understand the geometric structure of the physical region $\Omega$, e.g. its facial structure \cite{DESA1994451, Saunderson}. It will shed a light onto the distribution of contractions responsible for the minimal number of additional neutrinos.

\section*{Acknowlegments}
We would like to thank Marek Gluza for reading the manuscript and useful remarks. The work was supported partly by the Polish National Science Centre (NCN) under the Grant Agreement 2017/25/B/ST2/01987 and the COST (European Cooperation in Science and Technology) Action CA16201 PARTICLEFACE.

\section*{Appendix}
\addcontentsline{toc}{section}{\protect\numberline{}Appendix}%
In this appendix numerical procedure used to obtain results in Section \ref{sec3} is presented. Submatrices of the $T$-matrix in Tab.~\ref{tab:expbounds} will be denoted by t-matrix, i.e.  $\mathrm{t} \subseteq T$. Analysis has been done in two steps.

\begin{itemize}
\item[(i)] \underline{Step 1. Construction of lower triangular matrices.}

The lower triangular matrices are constructed  by the method introduced in subsection \ref{subsconstr}, with a given set of eigenvalues and singular values. 
In our approach only singular values are fixed and as eigenvalues coincide with diagonal elements, eigenvalues are randomly generated by exploring values of the $T$-matrix within the experimental ranges $ T_{11}, T_{22}, T_{33}$ (Tab.~\ref{tab:expbounds}), together with the requirement imposed by the majorization condition \eqref{eq:major}. Off-diagonal elements are adjusted by the appropriate rotations specified by eigenvalues and singular values using algorithm described in \cite{Kwong-Li:2001}.
We consider three scenarios 3+1, 3+2 and 3+3, which correspond to the subsets of the region $\Omega$ defined in (\ref{eqv1})-(\ref{eqv3}).
In each scenario and for different mass splitting cases (\texttt{I-III}), $10^8$ matrices were produced and compared with experimental bounds. This approach allows to find minimal "free" singular values smaller than one  (Section 3.3).
In this way we are able to see if it is possible to narrow down the experimental bounds on mixing matrix elements \eqref{tab:expbounds} resulting in new interval mixing matrix (Section 3.4).
However, as the analysis performed in this step is optimized for finding the minimal singular values, and not for investigating experimental bounds for mixing matrix elements, it is possible that the experimental mixing space is not fully covered. For example, in the 3+1 case, it may happen that there exist matrices from the $\Omega_{1}$ subset outside the obtained values but inside the experimental bounds.

To illustrate this issue, Fig.~\ref{fig2}  {shows} the resulting {t-matrix} entries for $\Omega_1$ and the mass scheme \texttt{(II)} (green cuboid). {In (\ref{eq:Tstep1})} numerical values for this matrix are given

 \begin{equation}  \mathrm{t}=
     \begin{pmatrix}
     0.97641\div0.99996 & 0 & 0\\
     0.00031\div0.02214 & 0.99331\div0.99999 & 0\\
     0.00048\div0.06892 & 0.00052\div0.01196 & 0.90040\div0.99985
     \end{pmatrix}.
     \label{eq:Tstep1}
 \end{equation}

In step 2 we improve the procedure in order to examine the whole range of experimental mixing limits.

\begin{figure}[h!]
\begin{center}
\includegraphics[scale=0.55]{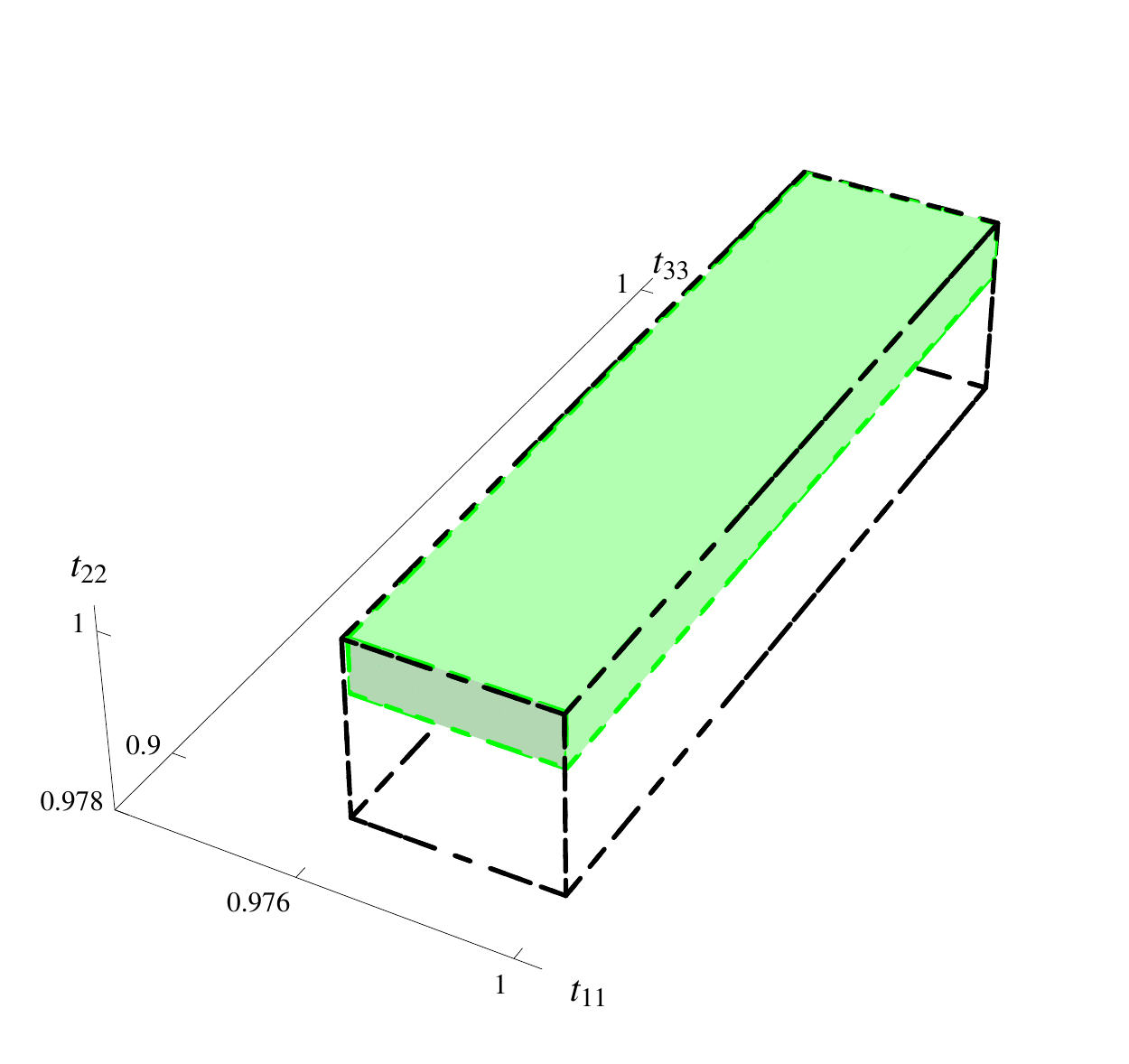}\includegraphics[scale=0.55]{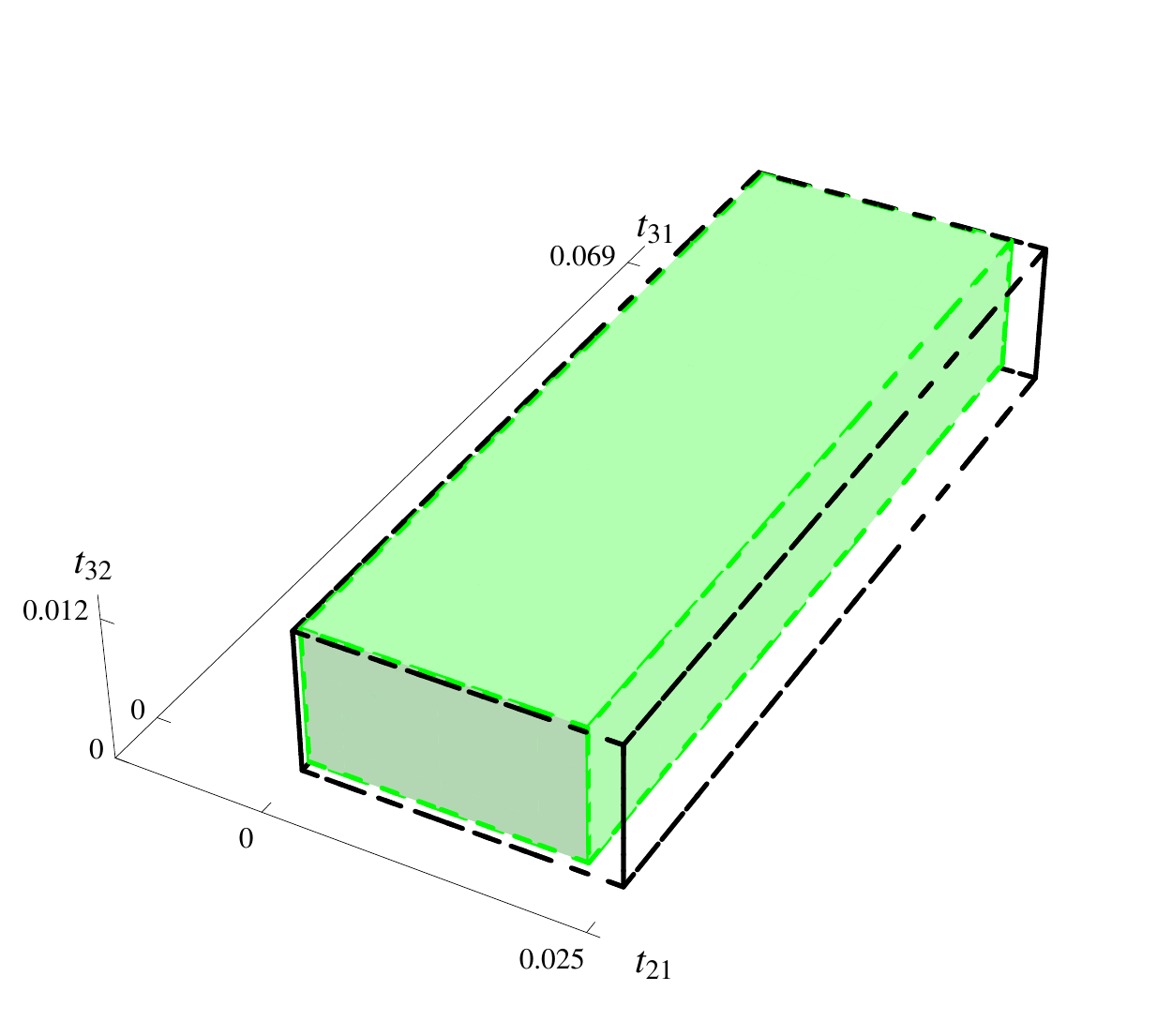}
\end{center}
\caption{Values of the  {t-matrix} obtained in the first step for the scenario with one additional neutrino and mass case \texttt{(II)} (green cuboid) compared with experimental limits (black-dashed cuboid). On left (right) the diagonal (non-diagonal) entries of the  {t-matrix} are shown.}
\label{fig2}
\end{figure}

\item[(ii)] \underline{Step 2. Sifting the boundary regions.}

If for any scenario the ranges of elements given in Tab.~\ref{tab:con} has shrunk by the procedure described in step 1, a single matrix element from the excluded region is fixed, and the remaining matrix elements are checked again within experimental bounds, inspecting strict division of $\Omega_1,\Omega_2,\Omega_3$ regions of $\Omega$ in (\ref{eqv1})-(\ref{eqv3}).
The procedure is repeated for the remaining values of the fixed matrix element from the excluded in step 1 region.  
As an example, {let us} take the {t-matrix} obtained in the previous step {for} the  3+1 scenario and the mass case \texttt{(II)} where two entries were {shrunk} significantly {($\mathrm{t}_{21}$, $\mathrm{t}_{22}$)}, see (\ref{eq:Tstep1}). We search for contractions from $\Omega_1$ subset in the excluded region starting with  {$\mathrm{t}_{21}=0.02215$}, checking the region within experimental limits for remaining entries:
 \begin{equation}  t=
     \begin{pmatrix}
     0.976\div1 & 0 & 0\\
     0.02215 & 0.978\div1 & 0\\
     0\div0.069 & 0\div0.012 & 0.900\div1
     \end{pmatrix}.
     \label{eq:Tstep2}
 \end{equation}
If any matrix from the (\ref{eq:Tstep2}) region is contraction in $\Omega_1$ subset, then the {$\mathrm{t}_{21}$} entry resulting from step 1 is no longer valid. In such case fixed value: $0.02215$ is the new upper limit for {$\mathrm{t}_{21}$} entry. Next we are increasing  {this} value and perform the same check. The step growth is 0.00001, and the upper experimental value is $T_{21}=0.02500$, see Tab.~\ref{tab:expbounds}. 
The procedure is repeated until all excluded region in step 1 is sifted. The same actions are performed for the rest of the lower-triangular entries of the {t-matrix}  (\ref{eq:Tstep1}).
This procedure allows to sweep systematically through the region of experimental matrix elements which were not covered by limits of the {t-matrix} obtained in step 1.

\end{itemize}


\bibliographystyle{elsarticle-num}
\bibliography{ref}

\end{document}